\documentclass[%
 reprint,
 amsmath,amssymb,
]{revtex4-2}

\usepackage{graphicx}
\usepackage{dcolumn}
\usepackage{bm}

\begin{document}

\preprint{APS/123-QED}

\title{Methane Sensing via Unbalanced Nonlinear Interferometry using a CMOS Camera}

\author{Jinghan Dong}
 \email{ip20409@bristol.ac.uk}

\author{Arthur C. Cardoso}%
\author{Haichen Zhou}%
\author{Jingrui Zhang}%
\author{Weijie Nie}%
\author{Alex S. Clark}%
\author{John G. Rarity}
\affiliation{ 
School of Electrical, Electronic and Mechanical Engineering, University of Bristol, Merchant Venturers Building, Woodland Road, Bristol BS8 1UB, UK}%

\date{\today}

\begin{abstract}
Here we present a high-sensitivity, rapid, and low-cost method for methane sensing based on a nonlinear interferometer. This method utilizes signal photons generated by stimulated parametric down-conversion (ST-PDC), enabling the use of a silicon detector to capture high-precision methane absorption spectra in the mid-infrared region. By controlling the system loss, we achieve more significant changes in visibility thereby increasing sensitivity. A low-cost CMOS camera is employed to capture spatial interference fringes, ensuring fast and efficient detection. The methane concentration within a gas cell is determined accurately. In addition, we demonstrate that ST-PDC enables long-distance sensing and the capability to measure open-path low ambient methane concentrations in the real world. 
\end{abstract}

\maketitle


\section{\label{sec:level1}Introduction}

Since 2014, the atmospheric concentration of methane has been increasing at double the rate of previous years\cite{fletcher2019rising}, exacerbating global warming as a potent greenhouse gas. Methane emissions are closely related to human activities, with primary sources including agriculture, energy production, and industry~\cite{KARAKURT201240}. Furthermore, in modern society, the frequent use of gas pipelines presents a common risk of methane leaks, which can become hazardous when mixed in certain proportions with air or other gases, potentially leading to explosions~\cite{PRENDESGERO2022104878}. Consequently, precise methane leak concentration measurements are essential, not only for ensuring safety but also to monitor emission sources promptly to mitigate global warming.

Optical methane sensors usually utilize the infrared absorption of methane, governed by the Beer-Lambert law. To avoid using expensive cooled detectors, absorption peaks between 1.65 $\mu$m and 1.66 $\mu$m are typically exploited. Tunable Diode Laser Spectroscopy (TDLS)~\cite{iseki2000portable} and Differential Absorption LIDAR (DIAL)~\cite{innocenti2017differential} are commonly used for remote direct sensing of methane, providing high accuracy and fast operation~\cite{TITCHENER2022118086}. Open-path Fourier Transform Spectroscopy (FTS)~\cite{griffith2018long} allows for the absorption detection of multiple gases across a broad spectral range. However, these methods's sensitivities are constrained by methane's weak near-infrared absorption peaks, which necessitate long interaction distances with the gas or low noise levels to ensure sufficient detection sensitivity.

According to the HITRAN database~\cite{gordon2022hitran2020}, methane has several absorption peaks in the mid-infrared (MIR) between 3.22 $\mu$m and 3.32 $\mu$m that are more than 60 times stronger than those around 1.66 $\mu$m. Theoretically, utilizing these MIR absorption peaks could yield higher sensitivity. Methods such as Cavity Ring-Down Spectroscopy (CRDS)~\cite{bahrini2015pulsed} and DIAL~\cite{VEERABUTHIRAN20151} can exploit these MIR peaks for direct sensing. However, detectors in the MIR typically require cryogenic cooling, are very expensive, have low detection efficiency and are noisy. To avoid using MIR detectors one can employ spontaneous parametric down-conversion (SPDC), a nonlinear process that generates correlated photon pairs with broadly spaced wavelengths, inside a nonlinear interferometer. This technique has been seeing increased use in sensing~\cite{lindner2023high}, imaging~\cite{lemos2014quantum}, microscopy~\cite{doi:10.1126/sciadv.abd0264, paterova2020hyperspectral}, optical coherence tomography~\cite{Paterova2018, Vanselow2020, Machado2020}, holography~\cite{Topfer2022, pearce2024single} and spectroscopy~\cite{Kalashnikov16, Tashima:24}. Quantum Fourier Transform Infrared (QFTIR) Spectroscopy~\cite{neves2024open, Lindner:21} utilizing SPDC has been employed for gas sensing. However, the resolution of gas absorption spectrum measurements using this technique is limited by the maximum optical path difference and the minimum wavelength step that can be scanned. Additionally, the low conversion efficiency of SPDC results in a low signal-to-noise ratio (SNR), which also impacts the sensitivity.

To address these problems, we use stimulated parametric down-conversion (ST-PDC), employing a continuous-wave (CW) pump laser and a CW MIR seeding source to generate the near-infrared wavelength signal light. This method allows us to take advantage of the strong absorption peaks of methane in the MIR while transferring the absorption feature onto near-infrared light that never interacted with the gas, which can be subsequently detected using a CMOS camera, avoiding the MIR thermal background~\cite{ma2023eliminating}. The stimulated light easily dominates the weak spontaneous light generated at the same pump power, the intensity of the signal generation is significantly higher than that generated by SPDC and has a narrow bandwidth, which results in an enhanced SNR (over SPDC based sensing). By changing the angle of the mirror that reflects the signal mode generated in the first pass, we can detect spatial interference fringes using a low-cost standard CMOS camera, enabling fast and high-resolution capture of methane absorption spectra, and accurate determination of methane concentration. With this technique the system can measure the interference with no moving parts. By introducing artificial losses to the interferometer we can increase the sensitivity of the setup to sense very small methane concentrations~\cite{Gemmell2023, florez2022enhanced}. A long coherence length seeding laser enables long-range open-path detection and the capability to measure low methane concentrations in laboratory ambient air.

\begin{figure*}[t!]
\centering
\includegraphics[width=0.8\linewidth]
{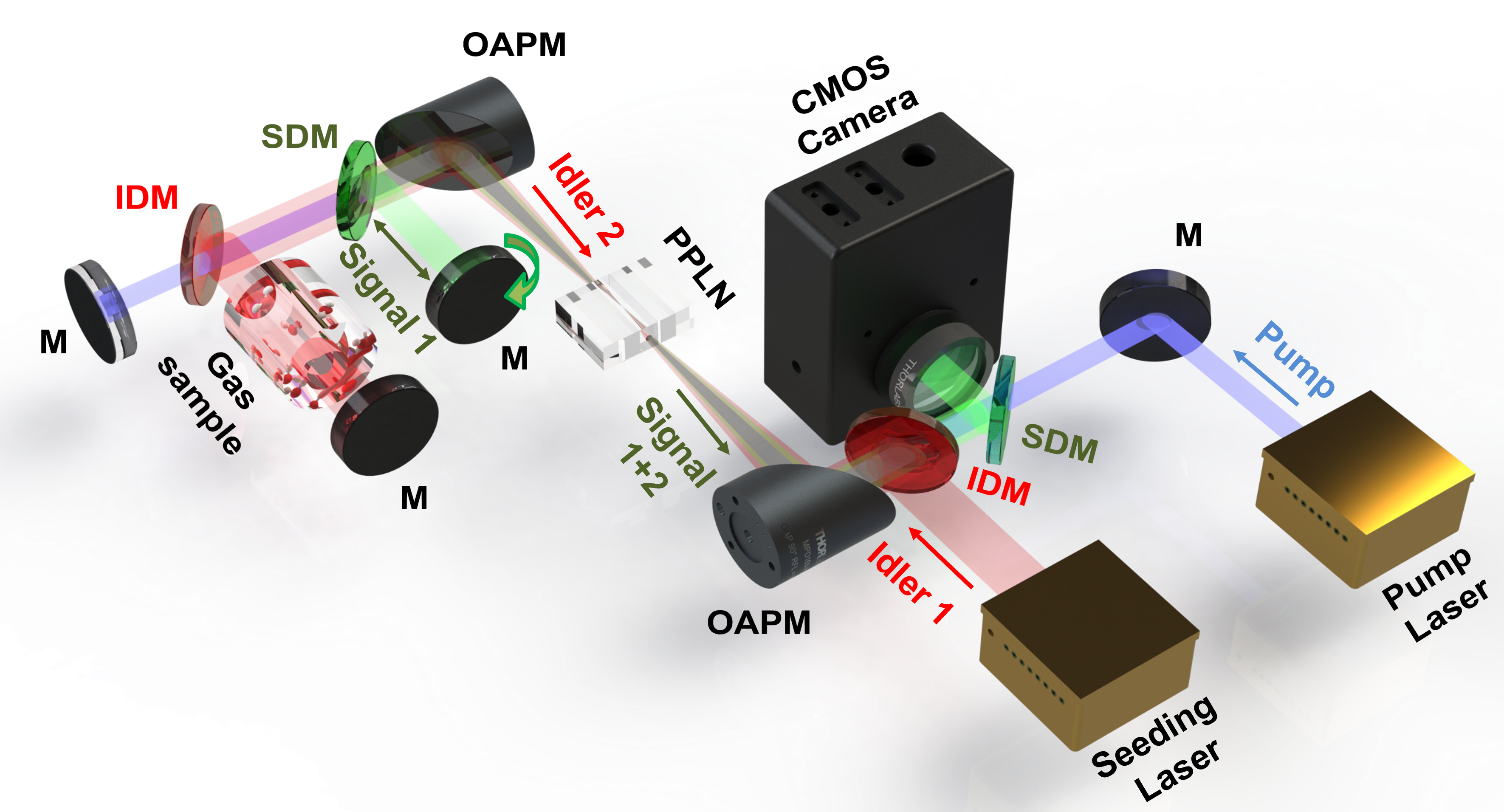}
\caption{Experimental setup. A 671 nm narrowband pump laser and a 3.22 $\mu$m narrowband tunable laser are focused by an off-axis parabolic mirror (OAPM) into the center of a periodically poled lithium niobate (PPLN) crystal. A signal beam (Signal~1) is generated via stimulated parametric down conversion (ST-PDC). Signal~1, the pump, and the seeding source are all collimated using a second OAPM. The idler, after passing through a methane gas sample, is reflected and seeds a second ST-PDC, generating a second signal mode (Signal~2) in the crystal with the reflected pump. Signal~1, which is reflected at a small angle by the mirror inside the interferometer, and Signal~2 create spatial fringes that are detected by a CMOS camera. M: mirror; SDM: signal dichroic mirror; IDM: idler dichroic mirror.}
\label{fig:experiment}
\end{figure*}

\section{Experiment}
Our experimental apparatus is illustrated in Fig.~\ref{fig:experiment}. The 671\,nm pump laser ($320 \, \text{mW}$, CW, narrowband, $>50\,m$ coherence length) and the 3.22\,$\mu$m tunable seeding laser ($0.6 \, \text{mW}$, CW, narrowband quantum cascade laser (QCL)) are both set to vertical polarization using half-wave plates and combined using a dichroic mirror. Then, they are coaxially aligned into an off-axis parabolic mirror (OAPM) with a focal length of $6$ inches. Due to the properties of the OAPM, the pump and idler can be focused at the same point in the center of crystal, generating the first 848\,nm signal mode, which we will call Signal~1. 

Here we use a $10\,\text{mm}$ long magnesium-oxide-doped periodically poled lithium niobate crystal (MgO:PPLN) designed for Type-0 phase matching. The crystal features a poling period of 16.6\,$\mu$m and is maintained at a temperature of $97.3^\circ\text{C}$ to maximize the generation of signal photons. Due to the narrow linewidth and long coherence length pump and seeding lasers, a narrowband signal can be generated. The Signal~1, idler, and pump beams are collimated by another identical OAPM. Signal~1 is then reflected by a long-pass dichroic mirror and reflected back by a slightly tilted mirror. After passing through the long-pass dichroic mirror, the idler is reflected by a 3.22\,$\mu$m short-pass dichroic mirror, then passes through a gas sample, and is retroreflected. The idler beam that interacted with the gas seeds the ST-PDC again, generating the second signal mode, Signal~2. The first and second signal generations both carry the intensity and phase information of the stimulating idler source.

By positioning the retro-reflecting mirror for the first signal generation at the focal point of the second OAPM, we can make an image of this mirror on a CMOS camera which is also placed at the focal point of the first OAPM. In this work, two signal beams are oriented at an angle horizontally, by slightly tilting the mirror for the first signal, producing vertical interference fringes. This enables the measurement of the whole interference pattern with a single shot on the camera, removing the need for a moving stage for the mirror, like a stepper motor or a piezoelectric actuator, to scan the interferometer phase and record fringes. The ratio of the intensities of the two signals affects the visibility of the interference fringes, which means the transmittance of the idler beam through the methane gas sample is reflected in the visibility. By adjusting the current of the seeding laser, we can change its wavelength to on-resonance or off-resonance with the methane absorption lines, observing a change in the visibility. This allows us to calculate the transmittance, thereby obtaining the methane absorption spectrum and calculating methane concentration. Detailed calculations related to this process will be discussed in the next section.

\section{Results}
\subsection{\label{sec:level2} Methane sample sensing}

In this work, we determine the methane absorption spectrum and concentration by analyzing the visibilities of spatial fringes of the signal mode, which enables the calculation of the transmittance across the methane absorption wavelengths (on-resonance) and transparent wavelengths (off-resonance) in the MIR region. The expression for visibility is~\cite{cardoso2024methane}.

\begin{align}
V = \frac{2 \sqrt{\alpha} \tau}{1 + \alpha \tau^2}\text{,} \label{eq:visibility} \tag{1}
\end{align}
where $\alpha$ is proportional to the ratio of Signal~1 and Signal~2 at the camera, and hence depends on signal, idler and pump losses in the system, while $\tau$ denotes the single-pass transmittance of the gas.

Assuming the off-resonance transmittance is unity allows us to measure $\alpha$. By tuning the wavelength of the seeding laser, which affects the transmittance of light through methane, we can use a CMOS camera to capture a series of spatial fringes with different visibilities produced by the two signals. It can be seen from Eq.~\ref{eq:visibility} that the change in visibility is not a linear relationship with the change in transmittance. Therefore, to improve measurement sensitivity in the presence of unavoidable technical noise, the loss of the system can be controlled to obtain a more pronounced change in visibility~\cite{florez2022enhanced}. Thus, it is necessary to analyze the optimal $\alpha$ theoretically.

\begin{figure}[t!]
\centering
\includegraphics[width=1\linewidth]{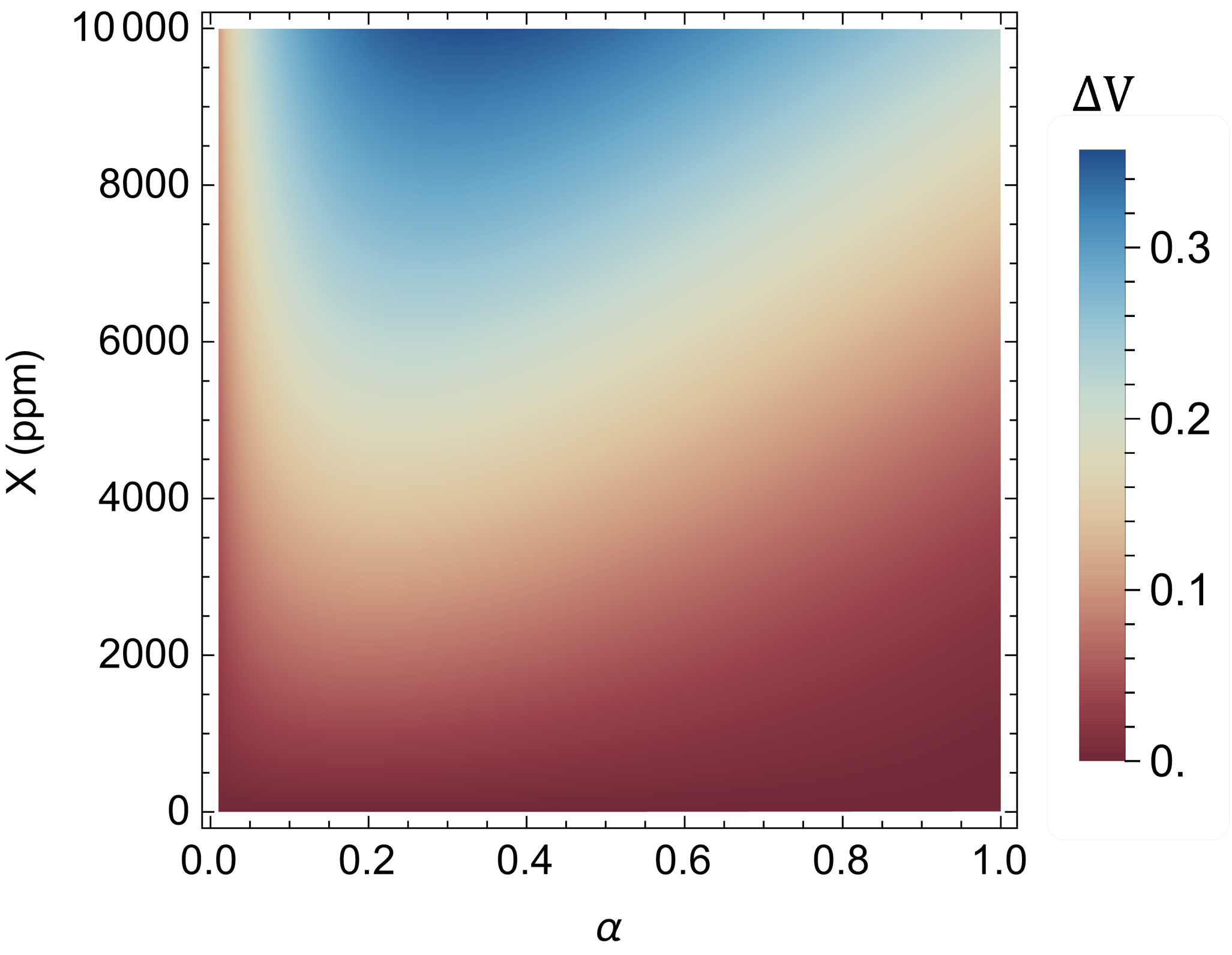}
\caption{ Simulation results of $\Delta V$ for different values of $\alpha$ and methane concentrations (in a 2.5 cm pathlength gas cell).}
\label{fig:2D_plot_X_deltaV_a}
\end{figure}

In this experiment, a gas cell of 2.5\,cm length, specified as 5000 ppm concentration by the manufacturer, was used. As shown in Fig.~\ref{fig:2D_plot_X_deltaV_a}, the visibility change $\Delta V$ for different concentrations of methane and values of $\alpha$ is illustrated. $\Delta V$ is the visibility difference at off-resonance and on-resonance wavelengths. By using the nominal concentration of the gas cell and its length, we can balance the interferometer to maximise $\Delta V$. Theoretically, for a concentration of 5000 ppm, this position is $\alpha \approx 0.25$, and the corresponding $\Delta V$ should be around 18.45\%.

\begin{figure*}[t!]
\centering
\includegraphics[width=1\linewidth]{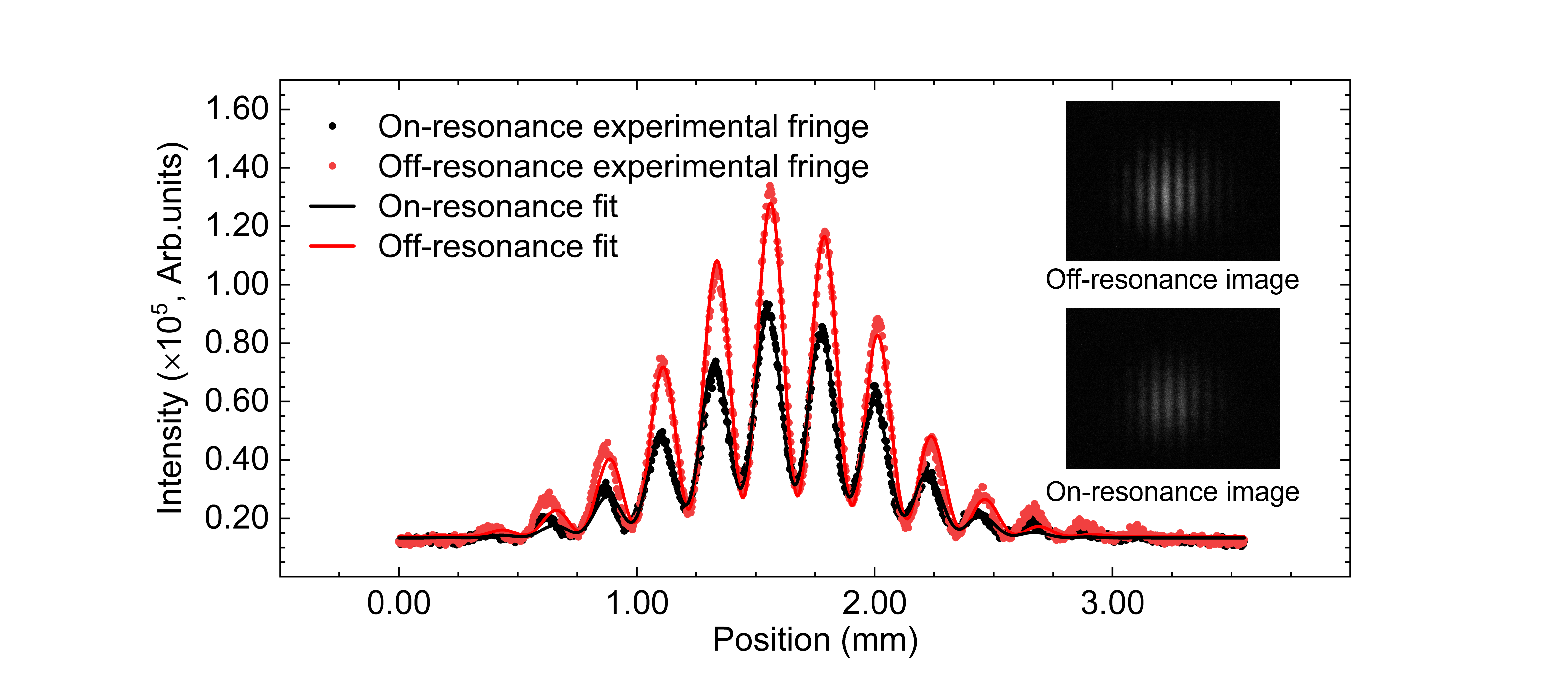}
\caption{ Sum of intensities for all pixels in each column as a function of the horizontal position of the pixels in the camera. The black and red dots with error bars represent the intensity distributions of spatial fringes at on-resonance and off-resonance wavelengths, respectively. The black and red lines represent the results of fitting these distributions with a Gaussian enveloped sine function (Eq.~\ref{gaussian interference}). The images at the top right and bottom right corners represent the spatial fringes captured by the camera at the off-resonance wavelength and on-resonance wavelength, respectively.}
\label{fig:short range fringes_with_images}
\end{figure*}

The images at the top right and bottom right corners in Fig.~\ref{fig:short range fringes_with_images} show the spatial fringes captured by the CMOS camera at the off-resonance and on-resonance wavelengths, respectively, with the exposure time set to 0.2~ms. The fringe is more visible at the off-resonance wavelength. As we adjusted the fringes to vertical, the intensity values of all pixels in each column captured by the camera can be summed, resulting in the intensity distribution shown in Fig.\ref{fig:short range fringes_with_images}. The x-axis represents the lateral position on the camera sensor (each pixel is 3.85~$\mu$m), and the vertical axis represents the total intensity of each column. The two signal beams are well overlapped at the camera, allowing us to use an interference model of two Gaussian beams without displacement for fitting. The model is described by the function
\begin{align}
I\left(x\right) = e^{-\left(\frac{(x - x_0)}{w}\right)^2} A \left(1 + V \cos\left(kx - b\right)\right) + C\text{\, ,}\label{gaussian interference}\tag{2}
\end{align}
where the Gaussian envelope is centered at \(x_0\) with the width of \(w\), the parameters \(A\), \(V\), \(k\), and \(b\) represent the amplitude, visibility, frequency, and initial phase of the interference fringes, respectively. The constant \(C\) accounts for the background noise of the camera. The visibility derived from this model contains information about the transmittance of the gas cell, that can be calculated using Eq.~\ref{eq:visibility}.

\begin{figure}[t!]
\centering
\includegraphics[width=1\linewidth]{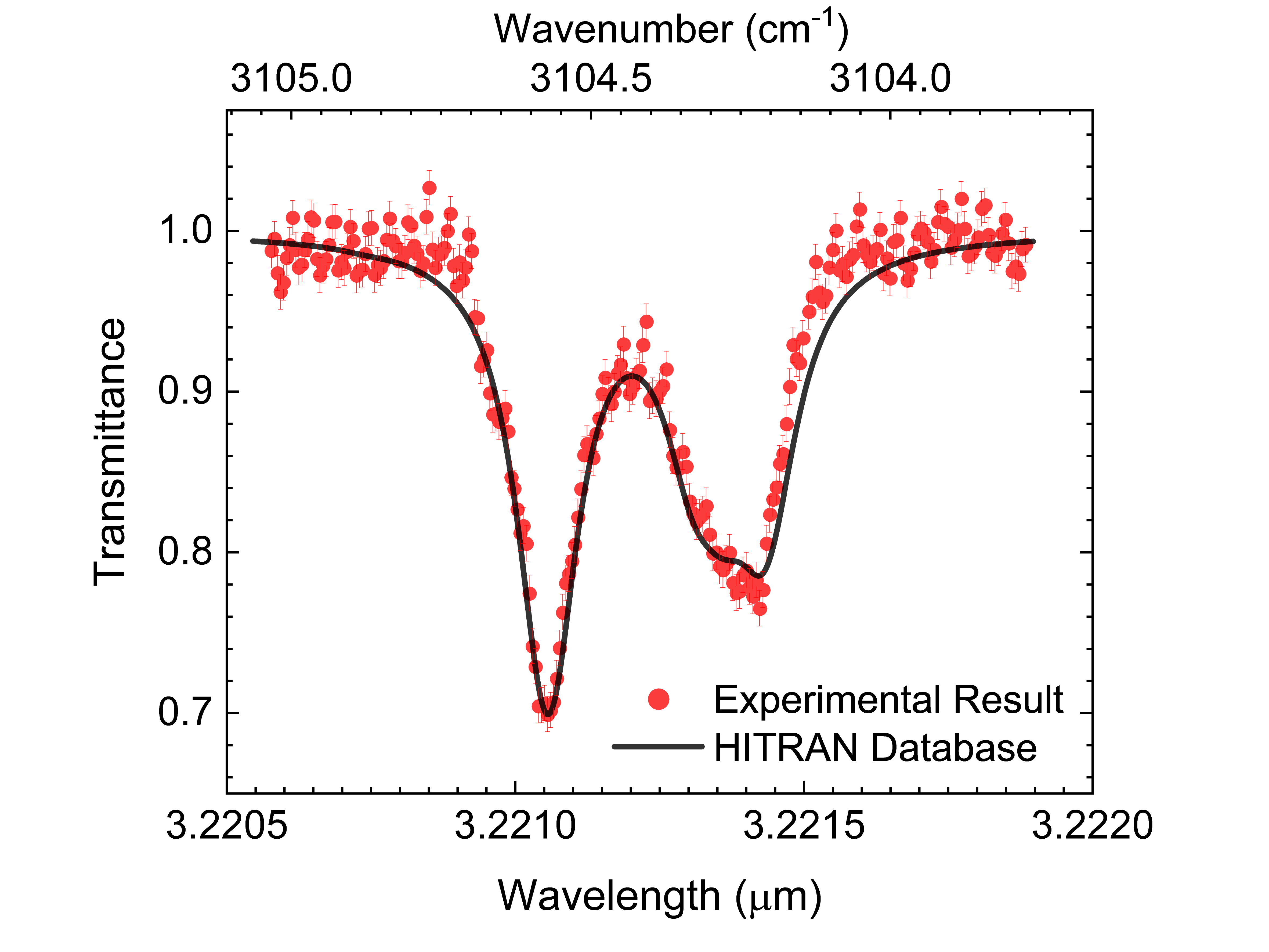}
\caption{ The absorption spectrum of a methane gas cell. The experimental transmittance data (red dots with error bars) and theoretical HITRAN data (black line) vary for different idler wavelengths.}
\label{fig:spectrum-short range}
\end{figure}

Figure~\ref{fig:spectrum-short range} illustrates the absorption spectrum of methane, with red dots with error bars denoting experimental data. The comparison with the black line theoretical data from the HITRAN database~\cite{gordon2022hitran2020} demonstrates that our method can achieve precise, high-resolution gas absorption spectra, capable of resolving methane dual absorption peaks separated by only 0.3~nm. The resolution shown in Fig.~\ref{fig:spectrum-short range} is the highest achievable resolution of $0.0056 \, \text{cm}^{-1}$. To clearly resolve two absorption peaks, a minimum scan time of $\sim1$ second is required, while achieving maximum resolution necessitates a duration of $\sim15$ seconds.

To accurately calculate methane concentrations, we initially employed the Voigt model, which is a convolution of Gaussian and Lorentzian profiles that incorporates Doppler and collision broadenings\cite{romanini2006optical} to precisely fit the methane absorption curve between 3.2205 $\mu$m and 3.2220 $\mu$m. This enabled the computation of a wavelength-specific absorption cross-section via the formula:
\begin{align}
X_{\text{CH}_4} = -\ln\left(\tau\right) / \left(Z n_{\text{air}} \sigma\right)\text{,} \tag{3}
\end{align}
where $Z$ is the path length of the idler passing through the gas, $n_{\text{air}}$ is the air molecule density, and $\sigma$ is the methane absorption cross-section. The concentration of methane is then determined using a weighted least-squares fitting of the experimental data, with each transmittance measurement error utilized as weights to optimize the fitting accuracy.

Based on the previous theoretical analysis, we used an 808\,nm half-wave plate as a 0.6-wave plate in the pump arm of the interferometer. After a double pass, the polarization was controlled to make the returning pump partially incompatible with the Type-0 phase matching condition. This approach can reduce the intensity of the second generation continuously, thereby balancing the system to obtain the required $\alpha$ value to detect the concentration of the gas cell. Adjusting $\alpha$ to $0.24 \pm 0.01$, the lowest point of $V_{\text{on}}$ was measured as $\left(61.7 \pm 1.0\right)\%$. The off-resonance visibility is $\left(79.1 \pm 0.9\right)\%$, averaged across all off-resonance wavelengths. The methane concentration in the gas sample fitted is $\left(4750 \pm 51\right)$ ppm. The uncertainty in fitting parameters, derived from the diagonal elements of the covariance matrix, reflects the standard errors of the model parameters.

\begin{figure}[t!]
\centering
\includegraphics[width=\linewidth]{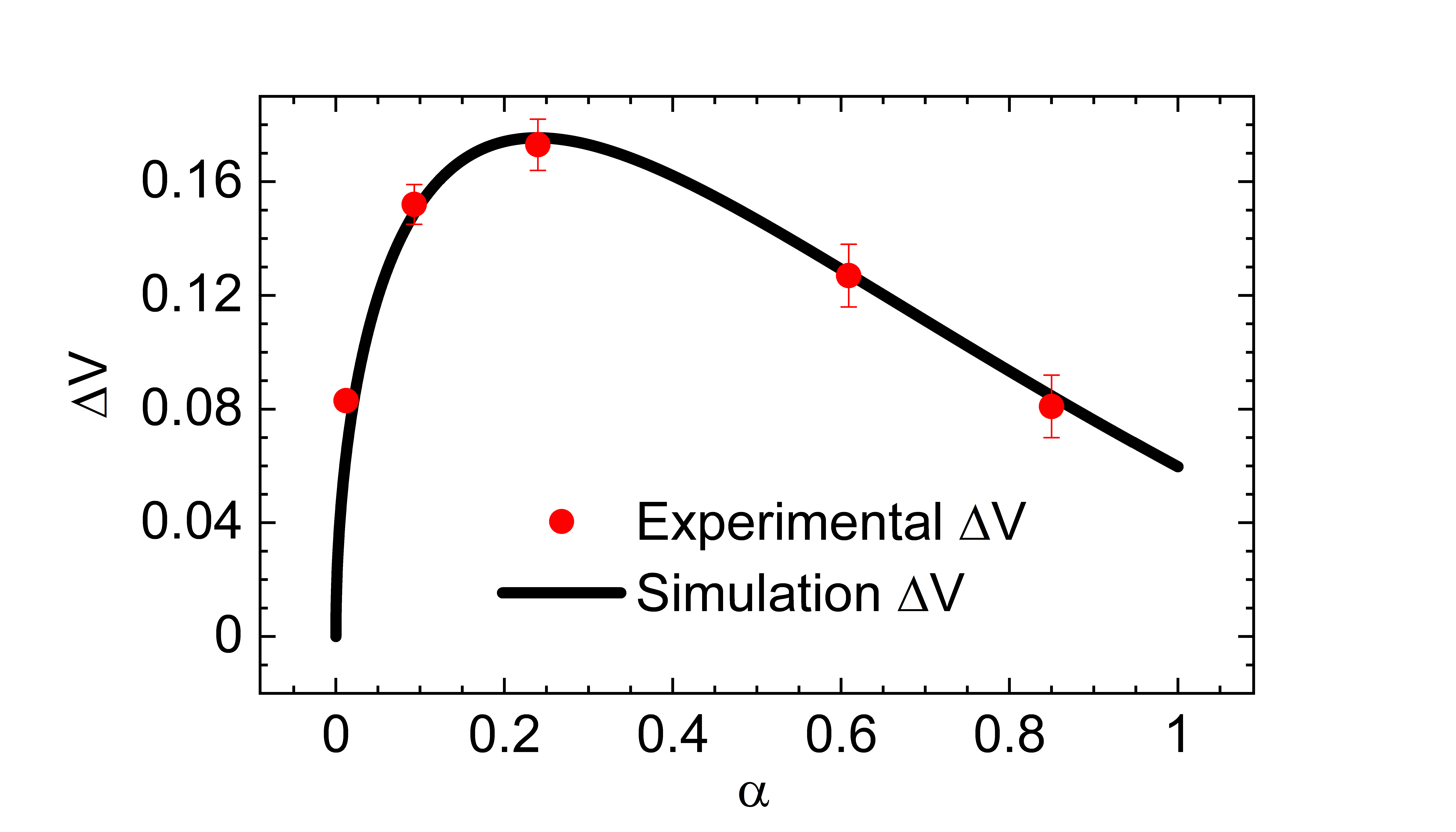}
\caption{ The relationship between visibility change ($\Delta V$) and system loss ($\alpha$) for a methane concentration of 4750 ppm. The red dots are experimental data, while the black line represents simulation results.}
\label{fig:deltaV_a}
\end{figure}

The relationship between $\alpha$ and the change in visibility at a concentration of 4750 ppm is depicted in Fig.~\ref{fig:deltaV_a}. The black line on the graph represents the theoretical curve calculated, and the red dots are the $\Delta V$ corresponding to different $\alpha$ values achieved by adjusting the polarization of the reflected pump in the experiment. It can be observed that the experimental results generally match the theoretical values. The third point at the top of the curve corresponds to the $\alpha$ value of $0.24 \pm 0.01$ used previously. This indicates that by pre-estimating the measurement range of concentration, an unbalanced interferometer can be used to maximize the measured $\Delta V$, thereby more clearly observing the change in visibility and enhancing sensitivity. Furthermore, $\Delta V$ varies slowly with respect to $\alpha$ near the peak, indicating that there is some flexibility in the experimental control of $\alpha$.

\begin{figure*}[t!]
\centering
\includegraphics[width=1\linewidth]{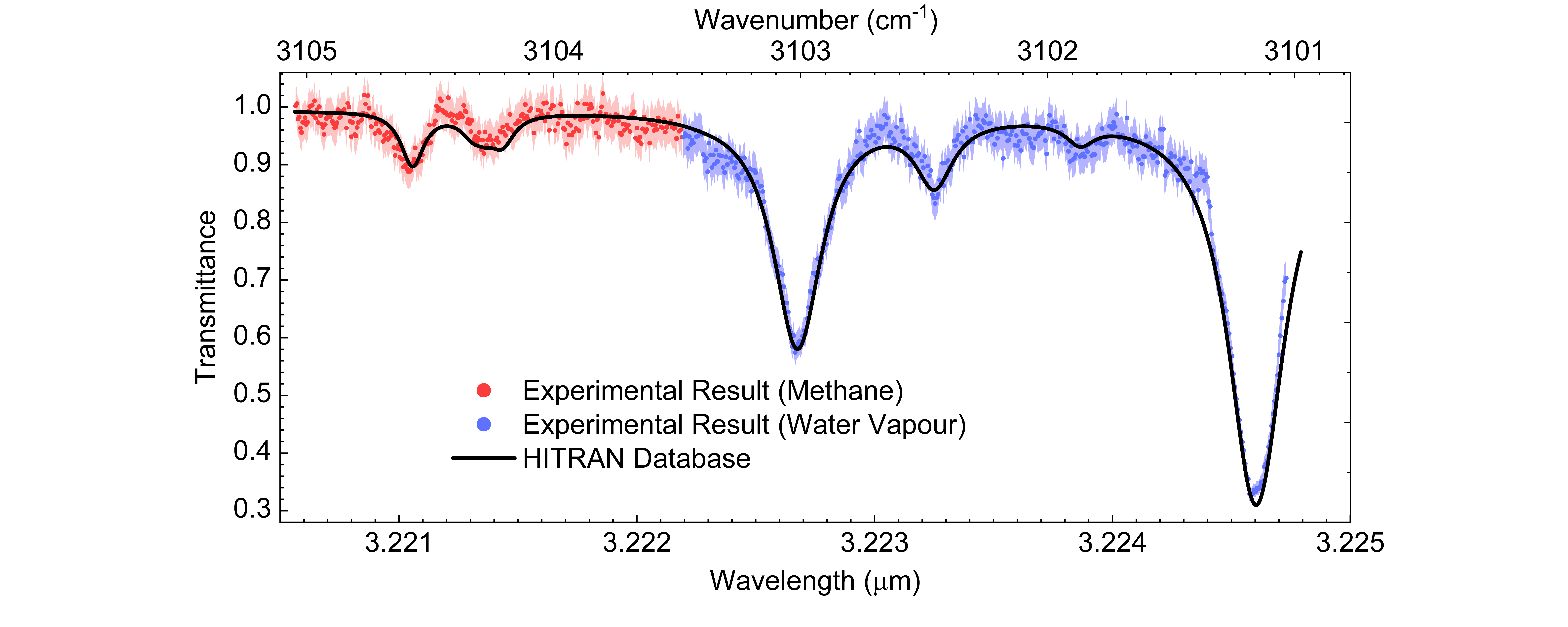}
\caption{ Absorption spectrum of methane (red dots with error band) and water vapor (blue dots with error band) in comparison with HITRAN data (black line) for 22 meters range in the lab.}
\label{fig:spectrum-long range}
\end{figure*}

\subsection{\label{sec:level2} Long-range open-path background gas sensing}

We have achieved long-range methane and water vapor sensing in ambient air between \(3.220 \, \mu\)m and \(3.225 \, \mu\)m. Due to the long coherence length of the MIR QCL laser used, spatial fringes can still be detected after removing the gas cell and moving the mirror reflecting the idler to a further distance. Based on the calculation of the Rayleigh length, we could extend the detection distance up to 100 meters. However, due to the limited space in the laboratory, we were only able to reach a maximum distance of 22 meters. Eight mirrors were used in the laboratory to fold the path to a total distance of 22 meters (11 meters to the final retroreflection mirror, and 11 meters back) for measuring methane concentration in the lab. Because of the inevitable loss over long distances, there is a 10 times reduction in the intensity of the second generation. Therefore, without sacrificing too much signal-to-noise ratio to achieve the highest $\Delta V$, we opted to measure under conditions where the first generation was attenuated by a factor of 4.

To accurately measure the low methane concentration at long distances, we faced several challenges. Due to the increased sensitivity to environmental vibrations over long distances, we have opted to increase the number of measurements to measure stable off-resonance and credible on-resonance visibility to detect $\Delta V$ caused by low methane concentrations. Twenty measurements were sufficient to obtain reliable results that can distinctly resolve the dual absorption peaks. By averaging the visibilities at each wavelength over multiple measurements, we mitigated the effects of random vibrations and phase drifts.  

According to the HITRAN database, the primary gases in the air with absorption peaks between \(3.220 \, \mu m\) and \(3.225 \, \mu m\) are methane and water vapor. The results of the spectra measurements at a 22~m distance are shown in Fig.~\ref{fig:spectrum-long range}, where the x-axis and y-axis represent the idler wavelength and average transmittance respectively. The black line in the figure is derived from HITRAN data, while the red dots with red error bands represent the methane absorption spectrum. Each point was measured 600 times with an exposure time of 0.1 ms, meaning a total of 0.06~s per point. The measured visibility off-resonance ($V_{\text{off}}$) is $\left(44.5 \pm 7.1\right)\%$, and at the point of the strongest absorption on-resonance ($V_{\text{on}}$) is $\left(39.5 \pm 6.9\right)\%$.
The methane concentration in the background air was determined to be $\left(3.09 \pm 0.14\right)$ ppm through fitting procedures. 

The blue dots and error bands in Fig.~\ref{fig:spectrum-long range} represent the absorption spectrum of water vapor. The visibility of the strongest absorption point of water is $\left(15.2 \pm 2.4\right)\%$. By employing the same fitting method used for methane, the absorption cross section for water vapor was obtained. Subsequently, the fitting gives the result of a water vapor concentration of $\left(9801.2 \pm 90.0\right)$ ppm. The laboratory temperature is $20^\circ$C. By calculating the ratio of the actual water vapor pressure and the saturation vapor pressure at this temperature, the relative humidity was found to be $\left(42.48 \pm 0.39\right)\%$, which is within the range $\left(40 \pm 5\right)\%$ measured by a commercial hygrometer.

\section{DISCUSSION}
In our previous work~\cite{cardoso2024methane}, a 1.55 $\mu$m wavelength signal was generated through ST-PDC using a 3.22 $\mu$m seeding source. By moving the position of one arm, interference fringes were detected with a single-photon detector. The precision of the subsequent methane concentration measurement was $10\%$, with a parametric gain of $10^{-8}$. We also theoretically analyzed the potential superior (shot noise limited) sensitivity of our method compared to direct sensing methods when the gain exceeded $10^{-4}$.
Here we have made significant improvements over this previous system. We have implemented the use of a standard CMOS camera to replace expensive single-photon detectors, shifting our signal detection wavelength to \(848 \, \text{nm}\) while still obtaining MIR methane absorption spectra. This substitution reduces cost, lowers technical noise, and increases efficiency and compactness~\cite{pearce2023practical}. By detecting spatial fringes with the camera instead of scanning the entire interference pattern by changing the optical path length, we can complete a transmittance measurement at each wavelength by waiting only for the camera exposure time of \(0.2 \, \text{ms}\). This significantly reduces the time required for concentration detection and completion of the entire spectral analysis. Furthermore, this increases the robustness of the interferometer as it is now free of moving parts. To achieve the same results as in our previous work~\cite{cardoso2024methane}, the scan time was reduced from tens of minutes to 3 seconds. 

Rapid detection also avoids influences from high-frequency vibrations, greatly improving the stability of measurements, allowing us to distinguish methane concentrations in the background air over long distances. Due to the use of higher power pump and the detection of shorter wavelength signals, along with a smaller beam waist in the center of the crystal, the gain factor has increased by more than $10^4$ times. When the seeding power is set at 30 $\mu$W, the power of the first-generation signal is 8 nW, measured by a photodiode power sensor. The resulting gain is calculated to be $2.7 \times 10^{-4}$. This enhancement means that under the same seeding power, a larger SNR is achieved, leading to increased sensitivity. Compared to the previous method that only uses two wavelengths to calculate the methane concentration and the corresponding errors, our newly proposed method to fit the entire absorption spectrum near \(3.22 \, \mu \text{m}\) provides more accurate and reliable concentration estimates. 

Compared to other optical methane detection methods, we have clear advantages. In contrast to direct detection methods~\cite{iseki2000portable,innocenti2017differential}, we utilize stronger methane absorption peaks and employ a silicon detector that is more efficient and cost-effective. In comparison with QFTIR spectroscopy~\cite{Tashima:24}, our method achieves over a hundred times higher resolution in gas absorption spectroscopy, and with the same pump power a higher SNR can be achieved. Furthermore, our technique does not require long scanning distances for higher spectral accuracy, which enhances system stability by eliminating the need for spatial movement. The long coherence length of the QCL used also enables us to perform long-distance gas detection.

\section{Conclusion }
This work demonstrates a methane sensor based on ST-PDC, capable of rapidly detecting gas concentrations and spectra at strong absorption peaks in the MIR region using a standard CMOS camera. We achieve a gas concentration detection accuracy of \(1\%\) for a gas sample and can sense ultra-low (background) methane and water vapor open-path concentration measurements at a distance of \(22 \, \text{m}\). The dynamic range of the sensor is increased by introducing losses and adjusting the parameter $\alpha$, allowing us to enhance sensitivity and measure low gas concentrations even in high-loss scenarios. Employing a CMOS camera to detect spatial fringes significantly shortens measurement time, completing the methane absorption spectrum from \(3.2205\) to \(3.2220 \, \mu\text{m}\) in 15 seconds with a resolution of 0.0056 $\text{cm}^{-1}$. Methane absorption peaks separated by \(0.3 \, \text{nm}\) can be distinguished. This method also has potential for applications in imaging and detection of other gases and substances.

\begin{acknowledgments}
We wish to acknowledge the support of QuantIC (EPSRC Project No: EP/T00097X/1), SPLICE (Innovate UK Project No: 106174) projects, and the China Scholarship Council (CSC)/University of Bristol joint-funded scholarship program. ASC acknowledges support from The Royal Society (URF/R/221019, RF/ERE/210098, RF/ERE/221060).
\end{acknowledgments}

\nocite{*}

\bibliography{apssamp}

\end{document}